\documentclass[aps,prl,final,twocolumn,superscriptaddress,floatfix,showpacs,10pt]{revtex4-1}

\usepackage[latin9]{inputenc}
\usepackage{graphicx,amssymb,soul,color,amsmath,bm}
\usepackage{epstopdf,hyperref}
\usepackage{times}

\hypersetup{colorlinks=true,citecolor=blue}

\sethlcolor{green}

\begin{document}

\title{
Quantum phase transition between orbital-selective Mott states in Hund's metals
}

\author{Juli\'an Rinc\'on}
\affiliation{Center for Nanophase Materials Sciences, Oak Ridge National Laboratory, Oak Ridge, Tennessee 37831, USA}
\affiliation{Materials Science and Technology Division, Oak Ridge National Laboratory, Oak Ridge, Tennessee 37831, USA}

\author{Adriana Moreo}
\affiliation{Materials Science and Technology Division, Oak Ridge National Laboratory, Oak Ridge, Tennessee 37831, USA}
\affiliation{Department of Physics and Astronomy, The University of Tennessee, Knoxville, Tennessee 37996, USA}

\author{Gonzalo Alvarez}
\affiliation{Center for Nanophase Materials Sciences, Oak Ridge National Laboratory, Oak Ridge, Tennessee 37831, USA}
\affiliation{Computer Science \& Mathematics Division, Oak Ridge National Laboratory, Oak Ridge, Tennessee 37831, USA}

\author{Elbio Dagotto}
\affiliation{Materials Science and Technology Division, Oak Ridge National Laboratory, Oak Ridge, Tennessee 37831, USA}
\affiliation{Department of Physics and Astronomy, The University of Tennessee, Knoxville, Tennessee 37996, USA}

\date{\today}

\begin{abstract}

We report a quantum phase transition between orbital-selective 
Mott states, with different localized orbitals, in a Hund's metals model.
Using the density matrix renormalization group, the phase diagram is constructed 
varying the electronic density and Hubbard $U$, at robust Hund's coupling. 
We demonstrate that this transition is preempted 
by charge fluctuations and the emergence of free spinless 
fermions, as opposed to the magnetically-driven Mott transition. 
The Luttinger correlation exponent is shown to have
a universal value in the strong-coupling phase, whereas 
it is interaction dependent at intermediate couplings. 
At weak coupling 
we find a second transition from a normal metal to the intermediate-coupling phase.
\end{abstract}


\maketitle

\textit{Introduction}.---Hund's metals (HM) are strongly interacting quantum states 
with bad metallic properties, where electronic correlations are dominated by the Hund's coupling $J$ 
and not by the Hubbard repulsion $U$.~HM are stable in intermediate coupling regimes, 
within two energy scales~\cite{Haule09,Yin12,Georges13}. A weak-coupling 
scale signals a transition from a bad to a coherent metal, i.e.~a Fermi liquid (FL). 
The second (strong-coupling) characteristic energy separates the incoherent metal 
from an ordered phase~\cite{Werner08,Haule09,Yin12,Georges13,Akhanjee13,Aron14}. HM display 
a variety of phenomena: mass enhancement~\cite{Haule09}, orbital selectivity~\cite{Anisimov02,deMedici11}, 
suppression of orbital fluctuations~\cite{deMedici11}, emergence of local moments~\cite{Werner08}, 
and non-Fermi liquid (NFL) physics~\cite{Haule09,Yin12,Georges13}.

Early work unveiled the freezing of local moments and power-law behavior in the electronic self-energy, 
using dynamical mean-field theory (DMFT) 
applied to a three-orbital Hubbard model~\cite{Werner08}.~A related orbital-dependent 
power law in the self-energy and optical conductivity was reported in DFT+DMFT studies~\cite{Yin12}. 
These results were discussed in the context of iron-chalcogenide and ruthenate superconductors 
suggesting that these materials are governed by Hund's physics~\cite{Vojta10,Lanata13,Hardy13,Georges13}. 
Using effective low-energy Kondo Hamiltonians with orbital degrees of freedom, 
the FL-NFL (coherent-incoherent) 
transition observed in those compounds was explained~\cite{Biermann05,Yin12,Akhanjee13,Aron14}.

The orbital-selective Mott phase (OSMP) is an example where Hund physics plays a crucial role. 
The OSMP is a bad metal where Mott insulator (MI) and normal metal coexist, 
leading to a NFL~\cite{Vojta10,Georges13}. This exotic behavior is due to the orbital-decoupling 
effect induced by $J$, suppressing orbital fluctuations. In the low-energy sector, the OSMP 
is described by a double-exchange model: a ferromagnetic Kondo lattice 
with band interactions~\cite{Biermann05,Vojta10,Rincon14}, which displays NFL. The conditions 
for a stable OSMP have been thoroughly discussed using several techniques~\cite{Liebsch04,Werner09,Ishida10,Liebsch11,deMedici11,deMedici11l,Bascones12,Georges13,Greger13}.

Studies of the magnetic order in the OSMP have shown a tendency to ordered phases 
such as paramagnetism (PM), ferromagnetism (FM), antiferromagnetism (AFM)~\cite{Chan09}, 
and block states (FM clusters coupled AFM). One of the properties of the OSMP state 
recently explored are its magnetic and charge orders: block states 
and FM were found in the spin sector and short-range order in the charge sector~\cite{Rincon14}. 

The OSMP has an associated orbital-selective Mott transition 
to a MI. This transition was first proposed to explain the coexistence 
of metallic and magnetic behavior in ruthenates~\cite{Anisimov02}. The origin of the 
orbital-selective Mott transition is mainly related to a strong $J$ and its band-decoupling effect~\cite{deMedici11}; 
however, crystal-field splitting~\cite{deMedici09} and unequal bandwidths are other 
factors that may lead to orbital differentiation and an orbital-selective Mott transition, in systems poorly 
hybridized. These findings were considered in the context of iron-based 
superconductors~\cite{deMedici09,Vojta10,Yu11,deMedici11,Yu12,Yi12,Luo13,Yu13,Georges13,Rincon14,deMedici14,Caron11,Caron12}.

In this paper we explore the influence of carrier doping and $U$ on an OSMP, employing 
a three-orbital Hubbard model and the density matrix renormalization group (DMRG)~\cite{dmrg1,dmrg2,dmrg3}. 
Our main result is the discovery of a formerly unknown quantum phase transition (QPT) between OSMPs (OSMP QPT) 
with different localized orbitals. We argue that this QPT is preceded by charge fluctuations and not magnetic fluctuations, and by the appearance of spinless fermions 
establishing a qualitative difference with the Mott transition. Calculations 
of the Luttinger liquid correlation exponent show a universal behavior 
in the strong-$U$ OSMP. We also find a small-$U$ QPT between a normal metal and an OSMP. 
This physics could be realized in 
heavy fermion and iron-based compounds with tendencies to Hund's metallicity.

\textit{Model and method}.---The model used in our study is a one-dimensional 
three-orbital Hubbard Hamiltonian. The details of the model were given elsewhere~\cite{suppl}, 
but here we briefly describe its main features. The Hamiltonian is 
divided as $H=H_{\rm kin}+H_{\rm int}$. The kinetic energy, $H_{\rm kin}$, includes 
hopping among orbitals $\gamma$ and an orbital-dependent 
crystal-field splitting term. The interacting part, $H_{\rm int}$, is composed of the intraorbital ($U$) 
and interorbital ($U'$) Coulomb repulsions, the FM Hund's coupling ($J$), and 
the pair-hopping term~\cite{Daghofer10}. 

The parameters in $H_{\rm kin}$ were chosen to mimic the band structure of 
the iron-based compounds~\cite{NoteBS}. We fix the ratio $J/U=1/4$, 
a prototypical value used in the study of multiorbital systems, and explore 
the phase diagram by changing the electronic filling in the range $3\leqslant n\leqslant 5$ 
and varying $U$. The total bandwidth is $W=2.45$~eV. Our conclusions are drawn 
from DMRG calculations of the orbital occupation number $n_\gamma$, the magnetic 
moment $\langle \mathbf{S}^2\rangle$, the charge $N(q)$ and spin $S(q)$ structure 
factors, and the orbital-dependent Luttinger liquid parameter $K_\gamma$. 
The truncation error was $<10^{-4}$; finite-size effects in the phase diagram are small~\cite{suppl}.

\begin{figure}
\centering
\includegraphics*[width=.8\columnwidth]{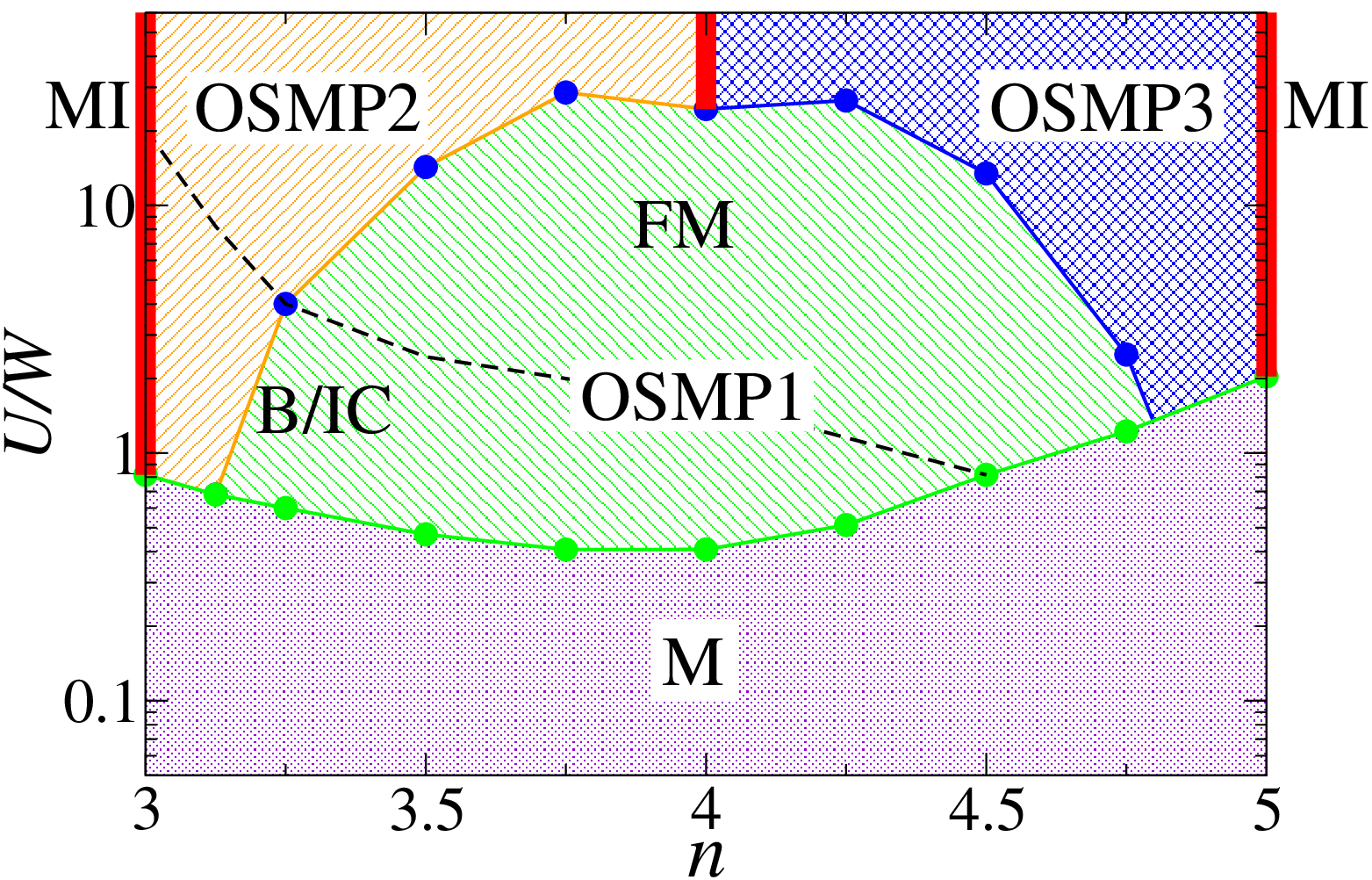}
\caption{DMRG phase diagram of the three-orbital model varying $U/W$ and $n$, at $J/U=1/4$ 
for a 72-orbital lattice. The phases are labeled as metal (M), Mott insulator (MI), and 
orbital-selective Mott phase (OSMP). In the OSMPs, 
block/incommensurate (B/IC) magnetism and ferromagnetism (FM) are separated 
by a dashed line. The quantum phase transitions are separated by the full lines.}
\label{fig:1}
\end{figure}

\textit{Results}.---We calculated the phase diagram of our multiorbital model varying 
$U$ and $n$~\cite{NoteMix}. The phase diagram is shown in Fig.~\ref{fig:1}. 
At integer $n$ and $U/W\gtrsim 1$, a MI with AFM is found (thick bars). 
For all values of $n$ and $U/W\lesssim 1$, a PM-M state is observed. Now one of the 
main results of our work: at intermediate and strong $U$, we detect 
phases with similar characteristics to those of the OSMP; specifically, three types 
of OSMPs are uncovered. ($i$) At intermediate $U$ we observe an OSMP with two 
metallic (itinerant) orbitals and \emph{one localized} orbital (OSMP1). ($ii$) At strong 
coupling and dependent on $n$, we unveil two new types of OSMPs. For $3<n<4$ the OSMP 
is characterized by the coexistence of one metallic orbital and \emph{two localized} orbitals (OSMP2). 
By contrast, for $4<n<5$ we observe an OSMP with one itinerant, \emph{one filled, 
and one localized} orbital, which translates to an effective two-orbital OSMP with one 
metallic and one MI orbitals (OSMP3).

The OSMP states display a nontrivial magnetic ordering: fully saturated FM 
is found in all of them, and block or incommensurate magnetism is seen 
in the OSMP1 and OSMP2 phases. Magnetic order was previously 
overlooked in mean-field studies of the OSMP; however, DMRG allows one to address 
issues of order. 
Moreover, though it 
is not explicitly shown, we have found regions of phase separation, where AFM and FM coexist, 
near $n=3$ and $5$. 
The magnetic orders reported in this work agree with those found in previous studies of the two-dimensional double-exchange model, which is the low-energy effective model of the OSMP~\cite{Millis95,Riera97,Aliaga01}, providing a connection with long-range ordered systems.

\begin{figure}
\centering
\includegraphics*[width=.8\columnwidth]{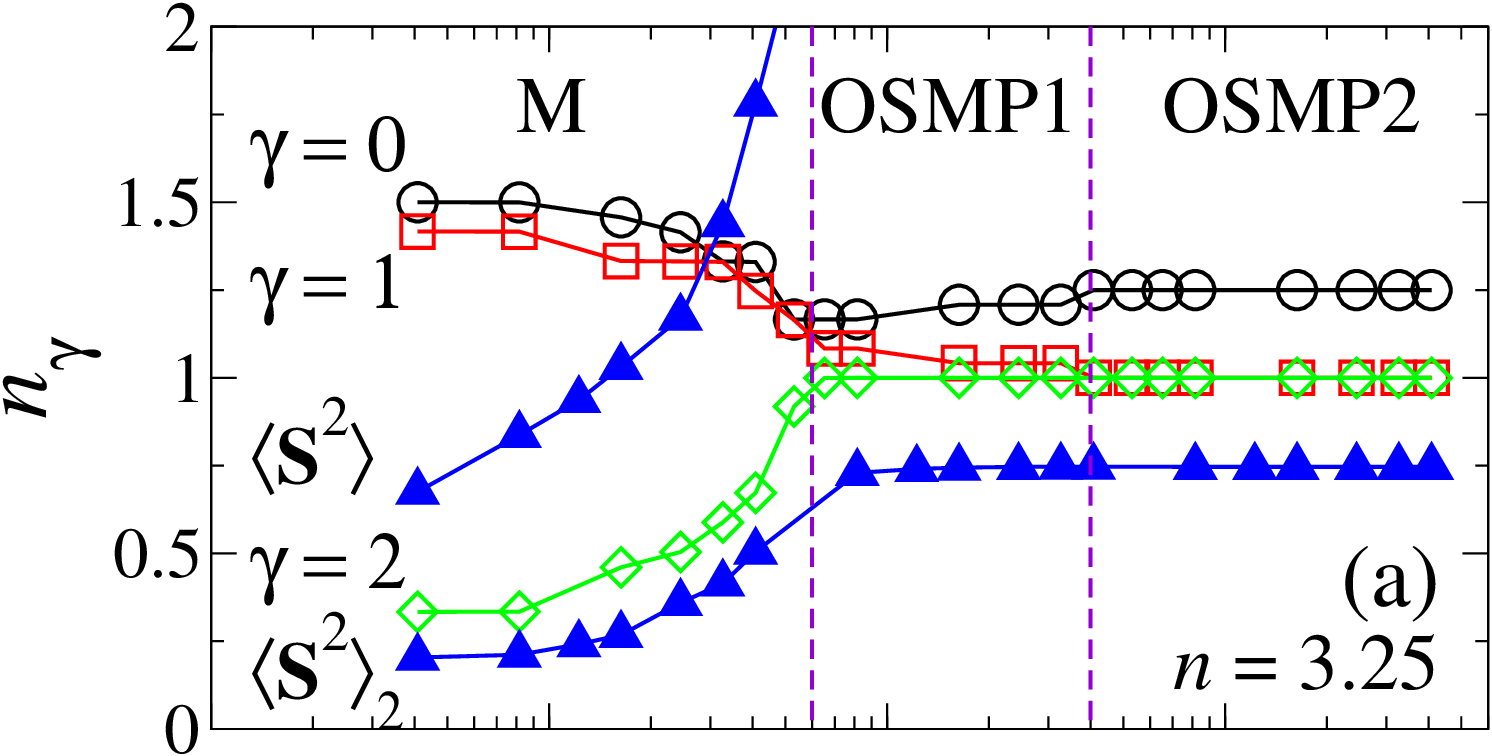}
\includegraphics*[width=.8\columnwidth]{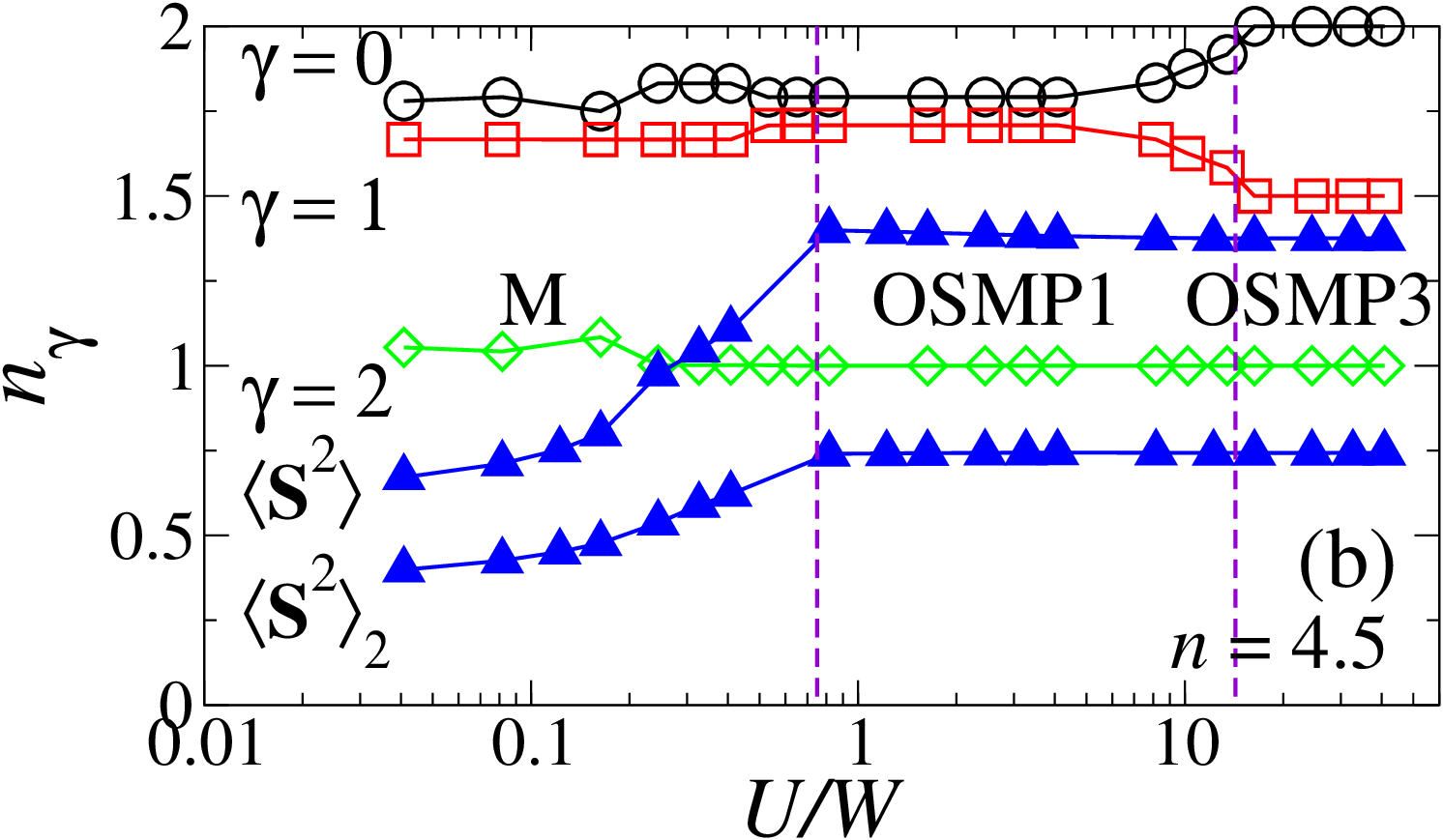}
\caption{Orbital occupation number, $n_\gamma$ (open symbols), and mean value 
of the total spin squared, $\langle\mathbf{S}^2\rangle$ (closed symbols), vs.~$U/W$, at $J/U=1/4$ for (a) $n=3.25$ and (b) $n=4.5$. $\langle\mathbf{S}^2\rangle_2$ is the magnetic moment 
for $\gamma=2$. The different phases are marked by vertical dashed lines. 
Notice the formation of a robust magnetic moment and a fairly $U$-independent 
$n_\gamma$ in the OSMP region.}
\label{fig:2}
\end{figure}

\begin{figure}
\centering
\includegraphics*[width=.9\columnwidth]{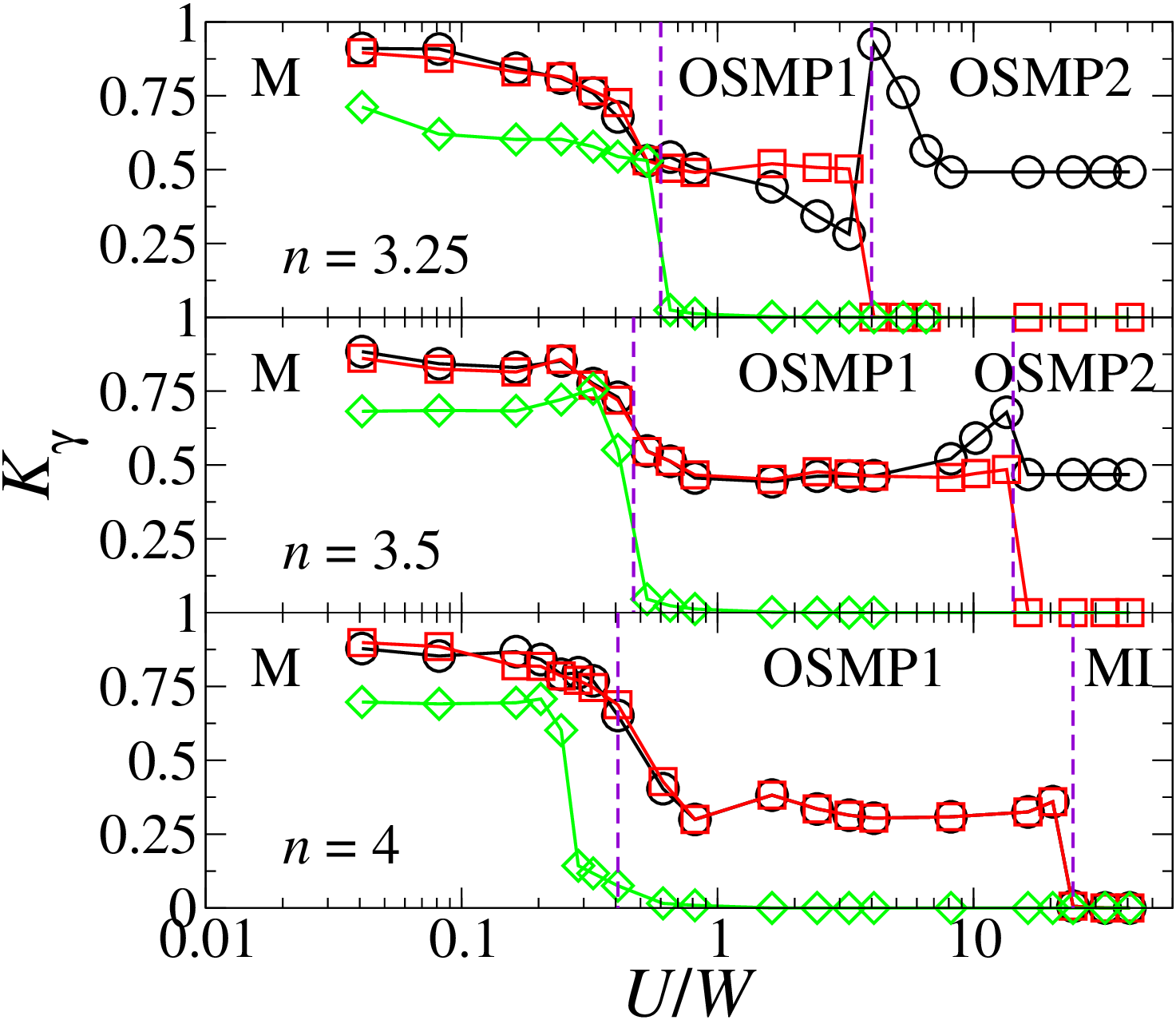}
\caption{Orbital-dependent Luttinger correlation exponent, $K_\gamma$, 
varying $U/W$, for several fillings $n$ and $J/U=1/4$. Note the existence of orbital-selective incompressible 
states ($K_\gamma=0$). The discontinuities correspond to quantum phase transitions.}
\label{fig:3}
\end{figure}

The emergence of different OSMPs can be deduced by monitoring 
the orbital occupation, $n_\gamma$, vs. $U/W$ (shown in Fig.~\ref{fig:2} 
for $n=3.25$ and $4.5$). In the small-$U$ region $n_\gamma$ exhibits a normal metal. 
As $U/W$ is increased $n_2$ locks to one, signaling the appearance of a MI in such 
orbital at intermediate coupling, while the other orbitals remain itinerant. 
This is the OSMP1 state shown in Fig.~\ref{fig:1}. Upon further increase of $U$ two 
different states develop depending on $n$. As shown in Fig.~\ref{fig:2}(a) 
at $n=3.25$, $n_1$ also localizes leaving only one orbital ($\gamma=0$) metallic; this 
is the OSMP2 state shown in Fig.~\ref{fig:1}. At $n=4.5$, plotted in Fig.~\ref{fig:2}(b), 
$n_1$ becomes doubly occupied, making it an inert orbital and inducing 
an OSMP with one localized and one itinerant orbital: this is the large-$U$ OSMP3 state 
shown in Fig.~\ref{fig:1}. One remarkable feature of the OSMP is that $n_\gamma$ 
is $U$ independent, but dependent on $J$, and strongly orbitally differentiated. 
The existence of the OSMP is further confirmed by the calculation of $K_\gamma$ (see Fig.~\ref{fig:3})~\cite{suppl}.

Figure~\ref{fig:2} also displays the magnetic moment 
$\langle \mathbf S^2\rangle$ vs.~$U$. In the metallic region we find 
a small $\langle \mathbf S^2\rangle$ as expected for a PM. 
In contrast, the OSMPs present a robust moment; this is true 
regardless of the value of the doping in the region $3<n<5$. The 
presence of a robust moment is a typical feature of the OSMP. 
The value of $\langle \mathbf S^2\rangle$ can be understood by considering 
the one-body contribution of each orbital. In the OSMP2, the occupations 
(in terms of holes because the system is above half-filling) are $n_0=0.75$, 
$n_1=n_2=1$ giving an effective $S_{\rm eff}\approx 1.375$ or $\langle \mathbf S^2\rangle=3.266$ 
which is fairly close to the actual value $3.3125$. The same 
argument holds for $n=4.5$, where $S_{\rm eff}\approx 1.313$ 
which is to be compared with $1.375$. These results agree 
nicely with the idea of the orbital-decoupling effect of $J$.

Assuming that the OSMP metallicity can be described by the universality class of the Luttinger liquid, we have calculated the correlation exponent, $K_\gamma$, which completely characterizes this theory. Figure~\ref{fig:3} shows the orbital-dependent $K_\gamma$ as a function of $U/W$ for several $n$. To avoid subtleties in the interpretation of $K_\gamma$, we have set the orbital hybridization to zero in $H_{\rm kin}$~\cite{suppl}. 
In the metallic region, we find that $K_\gamma$ is strongly orbital dependent (Fig.~\ref{fig:2}). 
This tendency is caused by the fact that by changing $U/W$ the filling $n_\gamma$ effectively 
changes as well, in a similar way as $K$ depends on $n$ for the single-orbital 
Hubbard model~\cite{Schulz90,GiamarchiBook}. By further increasing $U$, a critical 
line is crossed ($K_2=0$) signaling an incompressible state (MI); on the other hand, 
the metallic orbitals become more correlated: $K_\gamma^{\rm M} > K_\gamma^{\rm OSMP}$. 
With further increasing $U$, the second OSMP (for $n\neq 4$) or the MI (for $n=4$) is 
reached, where $K_1=K_2=0$, $K_1\rightarrow 1/2$ and $K_\gamma=0$, respectively. 
The presence of discontinuities in $K_\gamma$ are the result of different QPTs to different OSMPs.

Note that in the metallic region we found $1/2 < K_\gamma < 1$, resembling 
results for the standard Hubbard model~\cite{Schulz90,GiamarchiBook}. 
In contrast, in the strong-coupling OSMP $K_\gamma = 1/2$ or $K_\gamma=0$, 
which implies the onset of spinless fermions or an insulator, respectively (see below). 
One important conclusion can then be drawn from the behavior of $K_\gamma$: 
the strong-$U$ OSMP belongs to the universality class of \emph{free spinless fermions}. 
Calculating $K_\gamma$ with the DMRG is a challenging task because we need 
the long-range behavior of the correlations, which requires accurate calculations 
for large systems; therefore, the value of $K_\gamma$ calculated here 
should be considered as an upper limit to the thermodynamic value.

\begin{figure}
\centering
\includegraphics*[width=.8\columnwidth]{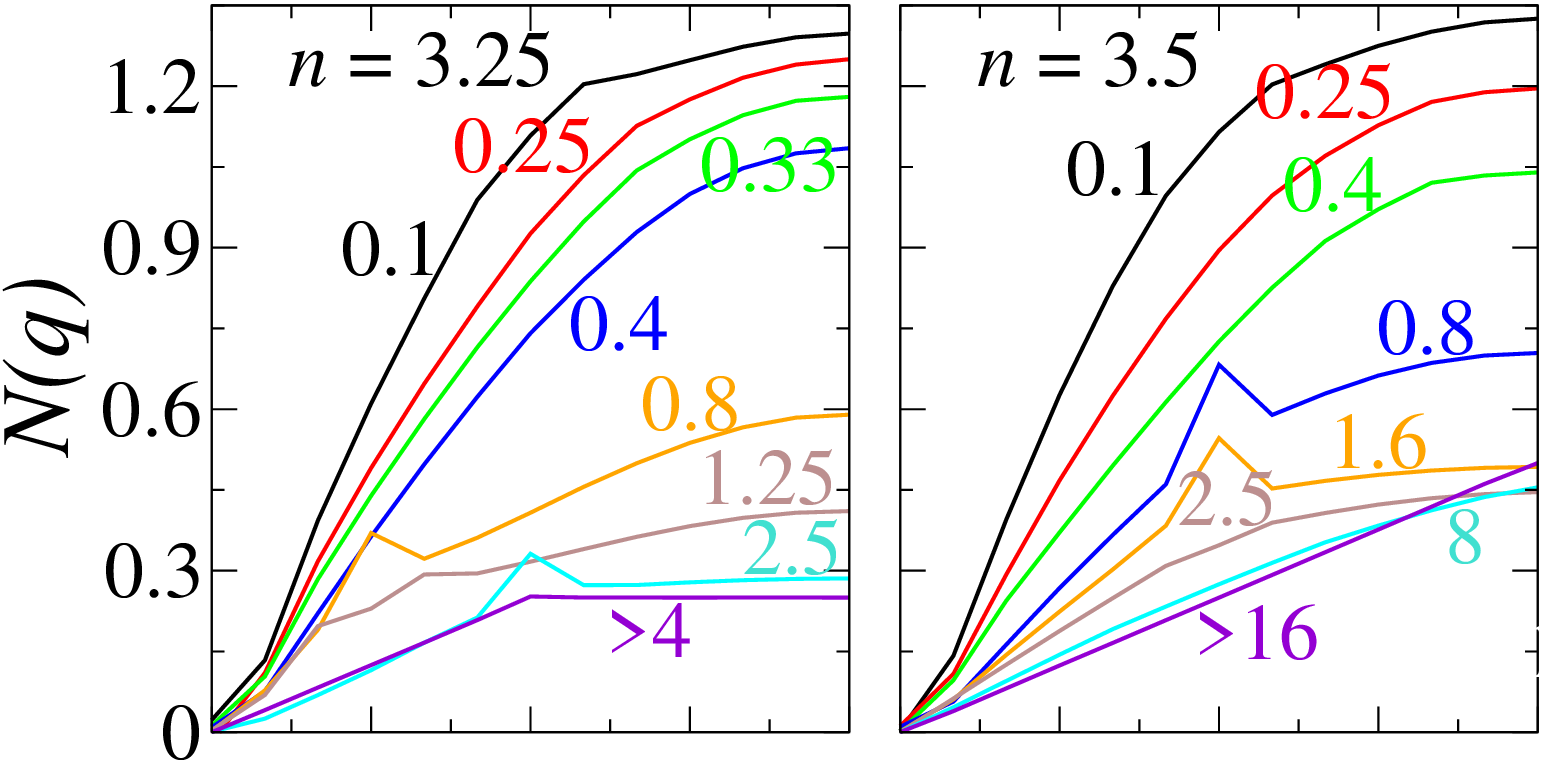}
\includegraphics*[width=.8\columnwidth]{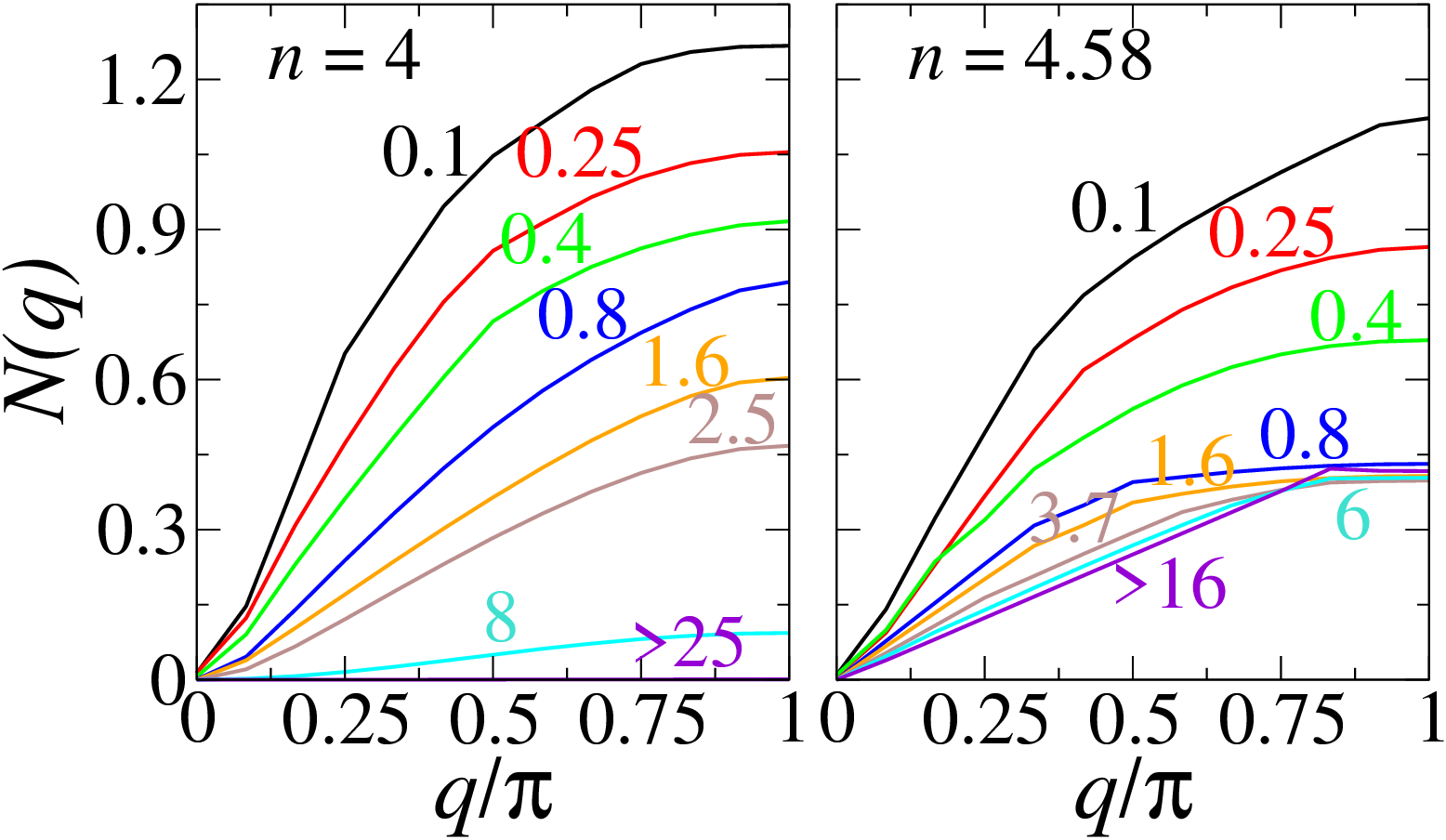}
\caption{Total-charge structure factor, $N(q)$, for several $n$ (panels) 
at $J/U=1/4$. Each panel shows $N(q)$ for different $U$ (color coded). Charge 
fluctuations are suppressed as $U/W$ increases, though not completely and depending 
on $n$. The quantum phase transition between orbital-selective 
Mott states is signaled by the onset of free spinless behavior with an 
$n$-dependent effective Fermi momentum.}
\label{fig:4}
\end{figure}

\begin{figure}
\centering
\includegraphics*[width=.82\columnwidth]{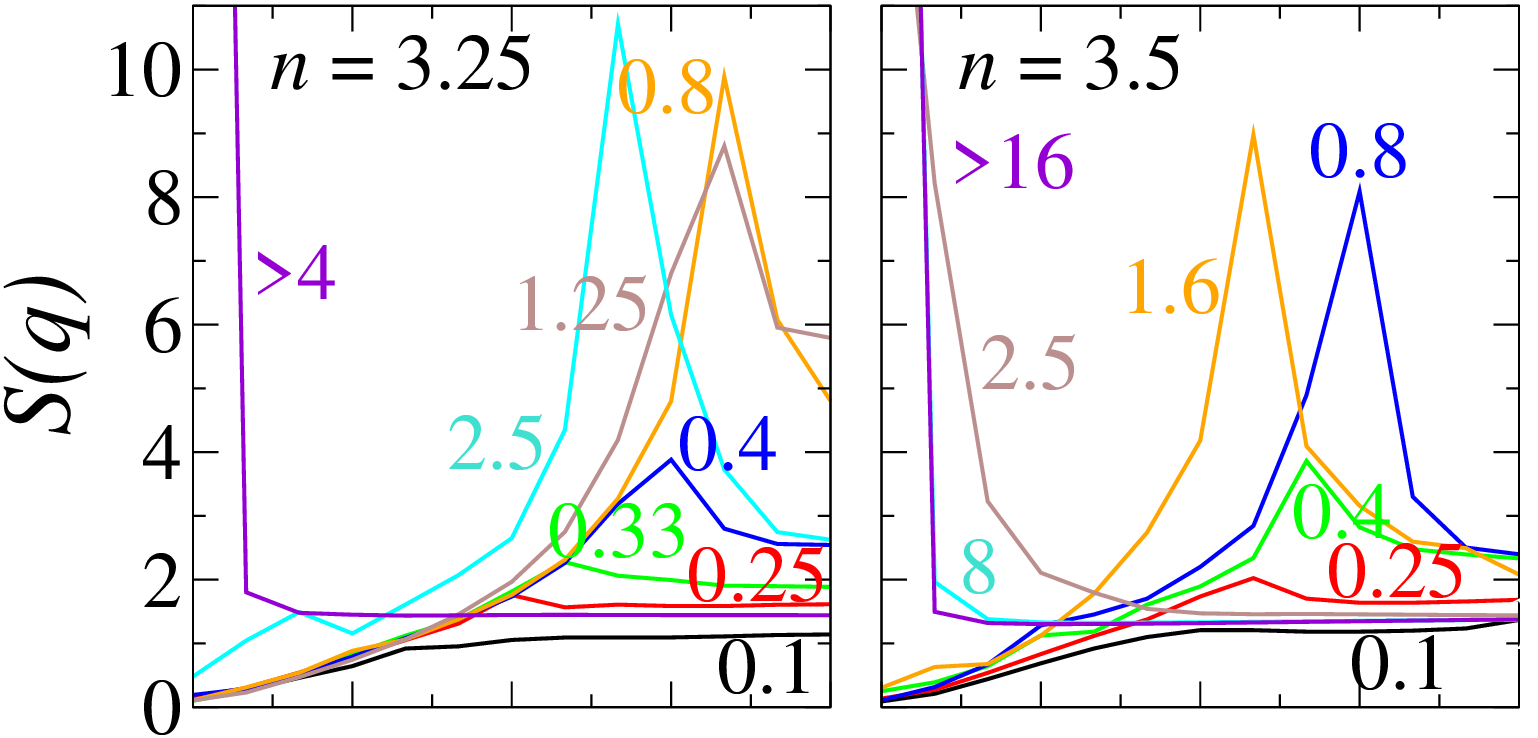}
\includegraphics*[width=.82\columnwidth]{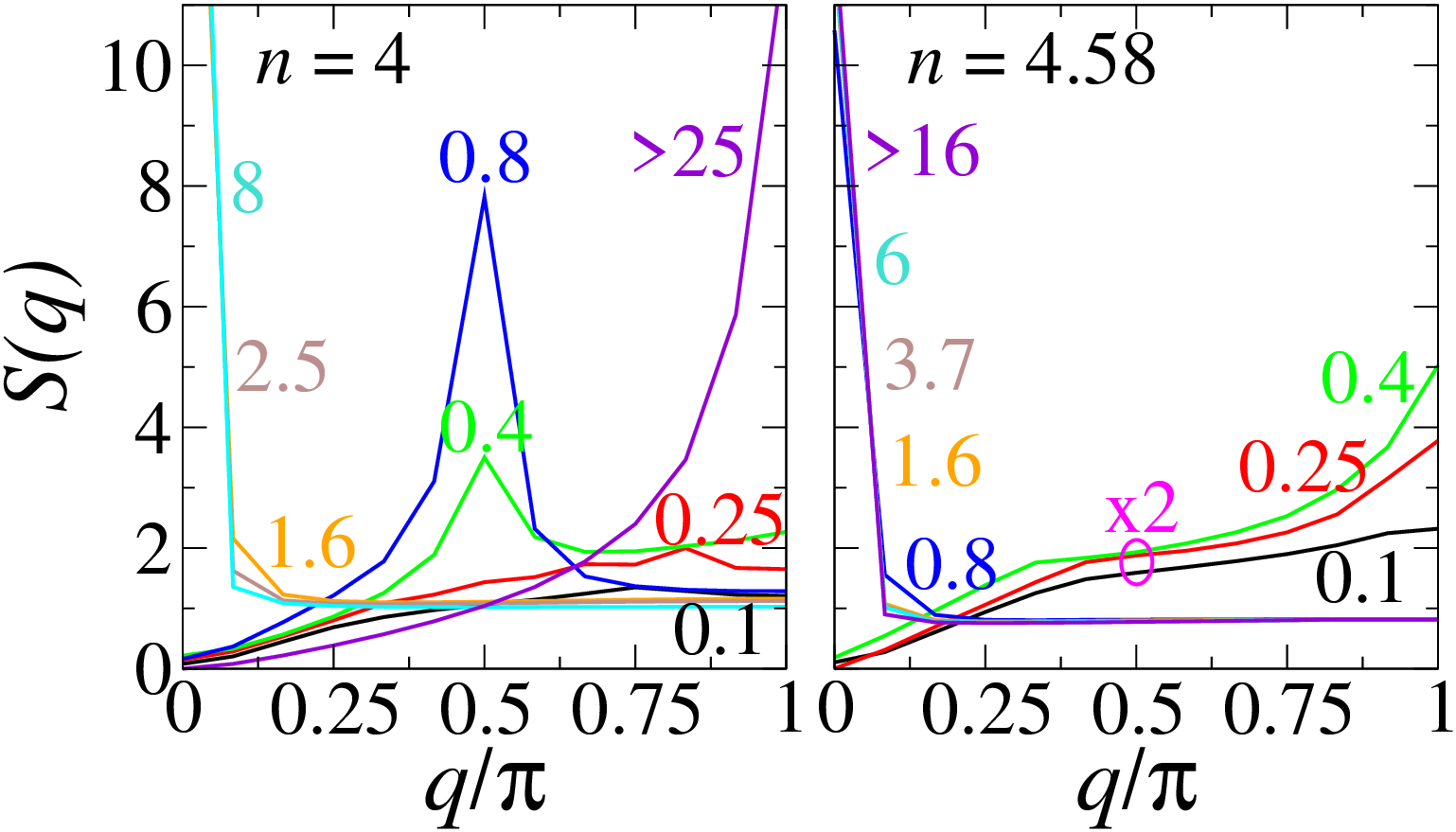}
\caption{Total-spin structure factor, $S(q)$, for the same parameters 
as in Fig.~\ref{fig:4}. Data sets are color coded with the corresponding 
value of $U/W$. Magnetic orders found are ferromagnetism (FM), 
block/incommensurate magnetism (B/IC), and for integer $n$ also 
antiferromagnetism (AFM). The magnitude fluctuations across the 
phase transition between orbital-selective Mott states remains 
largely unaffected showing that the transition is driven by charge fluctuations.
}
\label{fig:5}
\end{figure}

The origin of the OSMP QPT can be understood by monitoring 
the charge fluctuations across the transition. The total-charge 
structure factor $N(q)$, for different $n$ and $U$ (color-coded), 
is shown in Fig.~\ref{fig:4}. For \emph{noninteger} $n$, a general tendency is observed: 
as $U$ increases charge fluctuations are gradually suppressed until $U$ reaches 
the critical value of the OSMP QPT, and subsequently $N(q)$ takes the form of 
free spinless fermions. On the contrary, for $n=4$ where an orbital-selective Mott transition occurs, 
charge fluctuations are completely suppressed as $U\rightarrow\infty$. 
Unlike the Mott transition where charge fluctuations are 
almost completely frozen, in the novel OSMP QPT they still play 
a key role in describing the low-energy properties, where a correlated FM Kondo lattice 
arises as the effective model of the OSMP1, indicating entanglement between 
charge and spin~\cite{Biermann05,Rincon14}. 
It is important to remark that close to the OSMP QPT magnetic fluctuations 
are frozen, whereas for the orbital-selective Mott transition there is a change from FM to AFM (see Fig.~\ref{fig:5}).

The origin of the large-$U$ free-spinless-fermion behavior is different from that of the standard Hubbard model. For $U\rightarrow\infty$, all the electrons in each orbital, at a given site, will form unbreakable local triplets freezing charge fluctuations and effectively hopping as spinless electrons implying spin-charge separation and a drastic change of the screening properties 
of the low-energy FM Kondo lattice. The OSMP2 is different from the OSMP1 because the itinerant electrons 
do not effectively have spin and there are \emph{two} orbitals localized. 
Indeed Anderson and Hasegawa showed that for double-exchange models with $J\rightarrow\infty$ the resulting 
behavior is that of spinless particles~\cite{Anderson55}. Note that 
the effective Fermi momentum, $k_F^{\rm eff}$, of the spinless fermions 
will depend on $n$. The resulting $k_F^{\rm eff}$ can be extracted from 
the effective filling of the itinerant orbital of the strong-$U$ OSMP. 
For $n=3.25$, $\gamma=0$ is quarter-filled leading to $k_F^{\rm eff}=\pi/2$; 
for $n=3.5$, $\gamma=0$ is half-filled implying $k_F^{\rm eff}=\pi$ 
(see Fig.~\ref{fig:2}). However, incommensurate $k_F^{\rm eff}/\pi$ 
are also possible,  and Fig.~\ref{fig:4} shows this for $n=4.58$ which 
translates to $k_F^{\rm eff}\approx 0.84\pi$; notice that in this case 
it is $\gamma=1$ that is responsible for the spinless behavior, similar 
to $n=4.5$ shown in Fig.~\ref{fig:2}~\cite{suppl}.

Therefore, contrary to the orbital-selective Mott transition where the MI is preempted by AFM fluctuations, 
the OSMP QPT is preceded by the onset of charge order. Since the 
original model includes spinful electrons, the resulting spinless behavior 
corresponds to a nontrivial highly interacting state. To the best 
of our knowledge, such 
example of charge-fluctuation-enhanced Mott transition has never been reported.

Along with the OSMP QPT, we detected a small-$U$ stage from normal metal to OSMP. 
This QPT presents enhancement of charge and magnetic fluctuations (see
Figs.~\ref{fig:4} and~\ref{fig:5}). The enhancement manifests as 
peak changes close to the transition. A similar two-stage evolution 
from MI to OSMP to normal metal was reported in studies of  
chalcogenides~\cite{Yin12}.

The magnetic states found in Fig.~\ref{fig:1} are plotted 
in Fig.~\ref{fig:5}, for the same parameters as in Fig.~\ref{fig:4}. 
For $n=4$, we found FM, block, and AFM, as formerly reported~\cite{Rincon14}. 
For $n\neq 4$, we also found incommensurate magnetism which is characterized by peaks at fractional momenta in $S(q)$. 
As $U$ increases the general trend consists of moving from a small-$U$ PM to a block/incommensurate 
at intermediate $U$, to a large-$U$ FM ($n\neq 4$) or AFM ($n=4$). For the metal-OSMP 
QPT there is a change from PM to either block/incommensurate or FM. Thus, this transition is driven
by magnetic fluctuations. An entirely different 
situation is observed in the OSMP QPT, where $S(q)$ does not change across 
the transition. We notice that for $n\sim 4$, blocks of two- and three-site FM islands are found.

There is a caveat for the existence of the OSMP QPT: the presence of a FM state 
is necessary. Once FM is established the transition can occur. This can be seen by 
comparing $n=3.25$, where the transition happens simultaneously with an incommensurate-FM transition, 
and $n=3.5$ or $4.5$, where for an already existent FM state the transition occurs 
regardless of the magnetic fluctuations. Furthermore, the presence of phase separation 
around $n\gtrsim 3$ prevents its existence. 
The transition reported here 
is a generalized form of the orbital-selective Mott transition previously reported~\cite{deMedici11}.

We have confirmed our findings not only for $J/U=1/4$ but also for other ratios~\cite{suppl}. The specific critical $U_c/W$ required for the OSMP QPT are dependent on $J/U$. Note that these values are specific to a one-dimensional model with a specific hopping set. Our goal is to establish, via a generic example, that OSMP-OSMP transitions
are possible as a matter of principle, while specific realizations in real materials 
will surely require adjusting hoppings and likely dimensionality.

\textit{Conclusions}.---Using the DMRG, we report a QPT between OSMP states 
in a Hund's metals model. We have shown that this transition is signaled 
by furnishing of spinless-fermion behavior meaning that the transition is driven by charge 
and not magnetic fluctuations. The Luttinger liquid correlation exponent shows a universal value in the 
strong-$U$ OSMP and a weak interaction-dependent value at intermediate coupling. 
A small-$U$ transition from normal metal to OMSP is also found. These two transitions are similar 
to those found in realistic models for heavy-fermion and iron-based compounds.

J.R.~acknowledges insightful conversations with A.~Millis and K.~Al-Hassanieh. Support by the Early Career Research Program, U.S.~Department of Energy, (J.R., G.A.) is acknowledged. 
A.M.~and E.D.~were supported by the National Science Foundation under Grant No. DMR-1404375.


\begin{thebibliography}{10}
\bibitem{NoteBS} Our results are, however, quite general as the presence of the OSMP has been shown for different band structures modeling heavy-fermions, chalcogenides, and selenides. See Refs.~\onlinecite{Werner09,deMedici09,Yin12,Yu13,Rincon14}
\bibitem{NoteMix}
Notice that the phase boundaries will not qualitatively change if the hybridization is set to zero. This is particularly true for the OSMP, where the relevant energy scales are $U$ and $J$, and orbital decoupling occurs (see Refs.~\onlinecite{deMedici09,deMedici11}). The effect of $V$ can also be understood realizing that the low-energy theory of the OSMP is FM Kondo lattice, where such term is absent.
\end{thebibliography}
\end{document}


\title{
Supplemental Material for\\
Quantum phase transition between orbital-selective Mott states in Hund's metals
}

\author{Juli\'an Rinc\'on}
\affiliation{Center for Nanophase Materials Sciences, Oak Ridge National Laboratory, Oak Ridge, Tennessee 37831, USA}
\affiliation{Materials Science and Technology Division, Oak Ridge National Laboratory, Oak Ridge, Tennessee 37831, USA}

\author{Adriana Moreo}
\affiliation{Materials Science and Technology Division, Oak Ridge National Laboratory, Oak Ridge, Tennessee 37831, USA}
\affiliation{Department of Physics and Astronomy, The University of Tennessee, Knoxville, Tennessee 37996, USA}

\author{Gonzalo Alvarez}
\affiliation{Center for Nanophase Materials Sciences, Oak Ridge National Laboratory, Oak Ridge, Tennessee 37831, USA}
\affiliation{Computer Science \& Mathematics Division, Oak Ridge National Laboratory, Oak Ridge, Tennessee 37831, USA}

\author{Elbio Dagotto}
\affiliation{Materials Science and Technology Division, Oak Ridge National Laboratory, Oak Ridge, Tennessee 37831, USA}
\affiliation{Department of Physics and Astronomy, The University of Tennessee, Knoxville, Tennessee 37996, USA}

\date{\today}

\maketitle


In this supplemental section, technical details concerning the numerical calculation and the lattice model are provided, as well as additional results.

\section{Hamiltonian Model}
The Hamiltonian used in our calculations is a multiorbital Hubbard model composed of kinetic and interacting energy terms: $H=H_{\rm kin} + H_{\rm int}$. The kinetic part is
\begin{equation}
\begin{split}
H_{\rm kin} = &-\sum_{i,\sigma,\gamma,\gamma'} t_{\gamma\gamma'} \bigl( c^+_{i, \sigma, \gamma}c_{i+1, \sigma, \gamma'} + \mathrm{H.c.} \bigr) \\
&+ \sum_{i,\sigma,\gamma}\Delta_\gamma n_{i, \sigma, \gamma},
\end{split}
\end{equation}
where $t_{\gamma \gamma'}$ is a hopping matrix built in the orbital space $\{\gamma\}$ that connects sites $i$ and $i+1$ $(\gamma=0,\,1,\,2)$ of a one dimensional $L$-site system. The hopping matrix is:
\begin{equation}
t_{\gamma \gamma'} = 
\begin{pmatrix}
t_0 & 0 & -V \\
0 & t_1 & -V \\
-V & -V & t_2
\end{pmatrix}.
\end{equation}
In this matrix only hybridizations between orbitals 0 and 2, and 1 and 2, are considered. $\Delta_\gamma$ defines an orbital-dependent crystal-field splitting. The interacting term of $H$ is
\begin{equation}
\begin{split}
H_{\rm int} =& \; U\sum_{i,\gamma} n_{i,\uparrow,\gamma}n_{i,\downarrow,\gamma} 
-2J\sum_{i,\gamma<\gamma'} \mathbf{S}_{i,\gamma} \cdot \mathbf{S}_{i,\gamma'} \\
+& \left(U'-J/2\right)\sum_{i,\gamma<\gamma'} n_{i,\gamma}n_{i,\gamma'} \\
+&\, J\sum_{i,\gamma<\gamma'} \left( c^+_{i,\uparrow,\gamma}c^+_{i,\downarrow,\gamma} c_{i,\downarrow,\gamma'}c_{i,\uparrow,\gamma'} + \mathrm{H.c.} \right),
\end{split}
\end{equation}
where $U$ is the intra-orbital Hubbard repulsion and $J$ the Hund's rule coupling. The operator $c_{i,\sigma,\gamma}$ annihilates a particle with spin $\sigma$ in orbital $\gamma$ at site $i$, and $n_{i,\sigma,\gamma}$ is the particle operator at $i$ with quantum numbers $(\sigma,\gamma)$. 
The spin (density) operators are $\mathbf{S}_{i,\gamma}$ ($n_{i,\gamma}$) acting at site $i$ in orbital $\gamma$. In the case of SU(2) systems the following relation holds: $U'=U-2J$.

The ratio $J/U=1/4$ was fixed in all the calculations, and $U$ and the filling $n$ were varied. The values of the hoppings are (eV units): $t_{00} = t_{11} = -0.5$, $t_{22} = -0.15$, $t_{02} = t_{12} = V = 0.1$, $t_{01} = 0$, and $\Delta_0=0$, $\Delta_1=0.1$, $\Delta_2=0.8$; with a total bandwidth $W=4.9|t_{00}|$. Both hoppings and $W$ are comparable in magnitude to those used in more realistic pnictides models~\cite{Daghofer10}. These phenomenological parameters also reproduce qualitatively the position of the electron and hole pockets present in these materials.

\section{DMRG Details}
The density matrix renormalization group (DMRG)~\cite{dmrg1,dmrg2,dmrg3} results reported in this publication were obtained using open chains of lengths from 24 up to 150 orbitals, keeping up to 1300 states per block and up to 19 sweeps were performed in the finite-size algorithm. This choice of parameters gives a discarded weights in the range $10^{-6}-10^{-4}$. The accuracy for a poorer set of parameters was shown to be enough to get reliable observables with small finite-size dependence. Indeed, a similar behavior was previously observed in DMRG studies of multiorbital Hubbard models~\cite{Sakamoto02}. Part of this work was done with an open source code~\cite{dmrgpp}.

\section{Measurements}
For the sake of completeness, we show the explicit form of the observables calculated with the DMRG. The occupation number of each orbital is
\begin{equation}
n_\gamma = \frac{1}{L}\sum_{i,\sigma} \langle n_{i,\sigma,\gamma}\rangle.
\end{equation}

We have calculated correlation functions such as the total charge and spin correlations defined as follows: $\langle n_{i} n_{j} \rangle$, and $\langle \mathbf{S}_{i } \cdot \mathbf{S}_{j} \rangle$, where $n_i = \sum_{\gamma} n_{i, \gamma}$ and $\mathbf{S}_{i}= \sum_{\gamma} \mathbf{S}_{i, \gamma}$. In addition, we also calculated the orbital-dependent charge correlation functions $\langle n_{i,\gamma} n_{j,\gamma} \rangle$.

The structure factors for charge and spin are defined as
\begin{equation}
N_\gamma(q) = \frac{1}{L}\sum_{k,j} e^{-i q (j-k)} \langle (n_{k,\gamma} - n)(n_{j,\gamma} - n) \rangle,
\label{Nq}
\end{equation}
and
\begin{equation}
S(q) = \frac{1}{L}\sum_{k,j} e^{-i q (j-k)} \langle \mathbf{S}_{k} \cdot \mathbf{S}_{j} \rangle,
\label{Sq}
\end{equation}
respectively; a similar expression holds for $N(q)$. 

The calculation of the orbital-dependent Luttinger liquid correlation exponent, $K_\gamma$, was done by setting the hybridization to zero, $V=0$. The exponent is extracted by considering the limit of small wave vectors in $N_\gamma(q)$, $q\rightarrow 0$~\cite{Schulz90,GiamarchiBook}:
\begin{equation}
N_\gamma(q) \rightarrow K_\gamma q / \pi, \qquad q\rightarrow 0.
\end{equation}
Assuming that the structure factor behaves linearly, its slope is proportional to $K_\gamma$

\section{Additional Results}
\begin{figure}
\includegraphics*[width=.8\columnwidth]{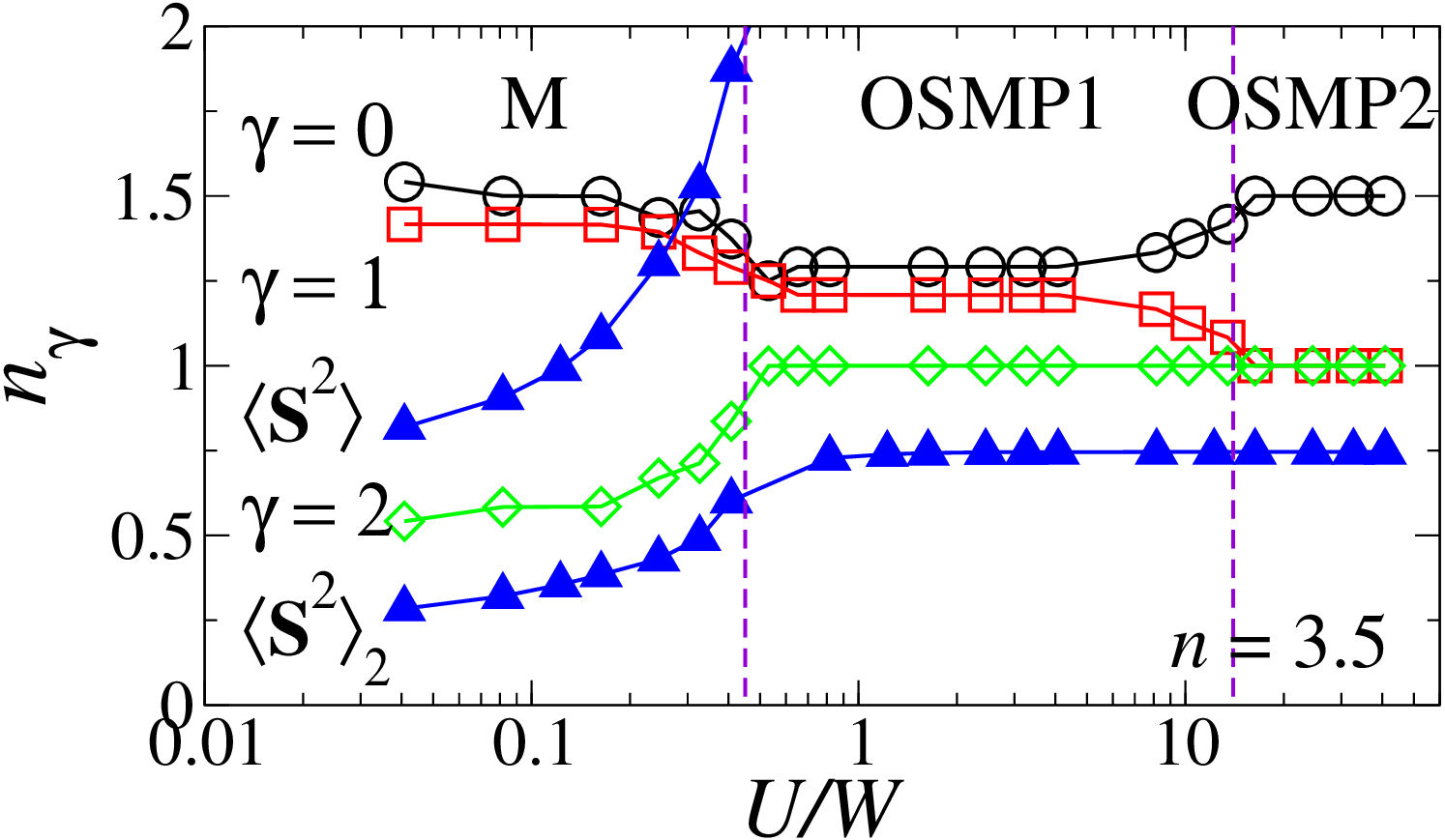}
\caption{Mean value of the orbital occupancy, $n_\gamma$ (open symbols), and mean value of the total spin, $\langle\mathbf{S}^2\rangle$ (closed symbols), vs.~$U/W$, at a fixed $J/U = 1/4$ and $n=3.5$. $\langle\mathbf{S}^2\rangle_2$ is the magnetic moment for $\gamma=2$. The different phases are marked by vertical dashed lines.}
\label{fig:S1}
\end{figure}

In this section, we present additional results on the quantum phase transition (QPT) between orbital-selective Mott phases (OSMP), with different degrees of localization. Figure~\ref{fig:S1} shows the orbital occupation $n_\gamma$ as a function of $U/W$ for filling $n=3.5$. As in the cases discussed in the main text, we observe the evolution from a metallic state for small $U$ towards an OSMP with one orbital localized and two itinerant ones, and then for the strong-$U$ limit, we see the advertized OSMP QPT, where the second phase corresponds to an OSMP with two orbitals localized and one itinerant. The formation of a robust magnetic moment within the OSMPs is a clear signature of its existence, which in this case reaches a value $\langle\mathbf{S}^2\rangle~\approx 2.875$ in the strong-$U$ regime. The evolution of $K_\gamma$ for this filling is shown in the main text in Fig.~3.

\begin{figure}[!b]
\includegraphics*[width=.8\columnwidth]{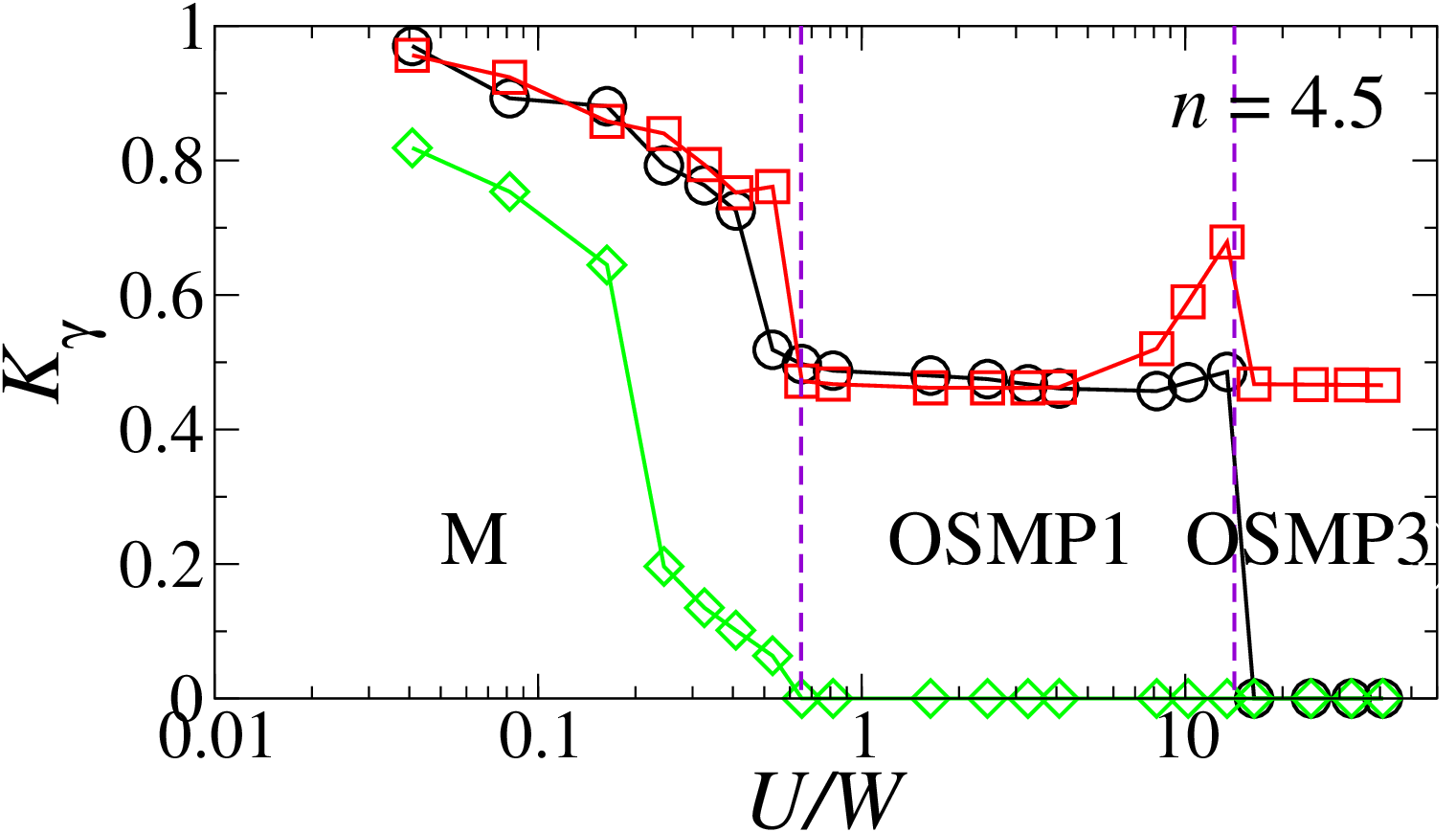}
\caption{Orbital-dependent Luttinger correlation exponent, $K_\gamma$, vs.~$U/W$, for $n=4.5$ and $J/U=1/4$. The abrupt changes correspond to quantum phase transitions (see Fig.~2(b) in main text).}
\label{fig:S2}
\end{figure}

In Fig.~\ref{fig:S2}, we plot the Luttinger correlation exponent for each orbital $\gamma$ versus the repulsion $U$ ($V=0$). The corresponding results for $n_\gamma$ are shown in the main text (see Fig.~2(b)). $K_\gamma$ exhibits a slightly correlated metallic behavior for $U/W < 1$. As the interaction increases the orbitals become more interacting as indicated by a decrease in $K_\gamma$. For $U/W\gtrsim 1$, orbital $\gamma=2$ localizes ($K_2=0$) while the itinerant ones have an exponent close to 1/2. Upon further increase of $U$, orbital $\gamma=1$ localizes as well, leaving a metallic orbital ($\gamma=0$) with $K_0=1/2$ signaling the onset of free-spinless-fermion behavior and the OSMP QPT (more details in the main text).

\begin{figure}
\includegraphics*[width=.8\columnwidth]{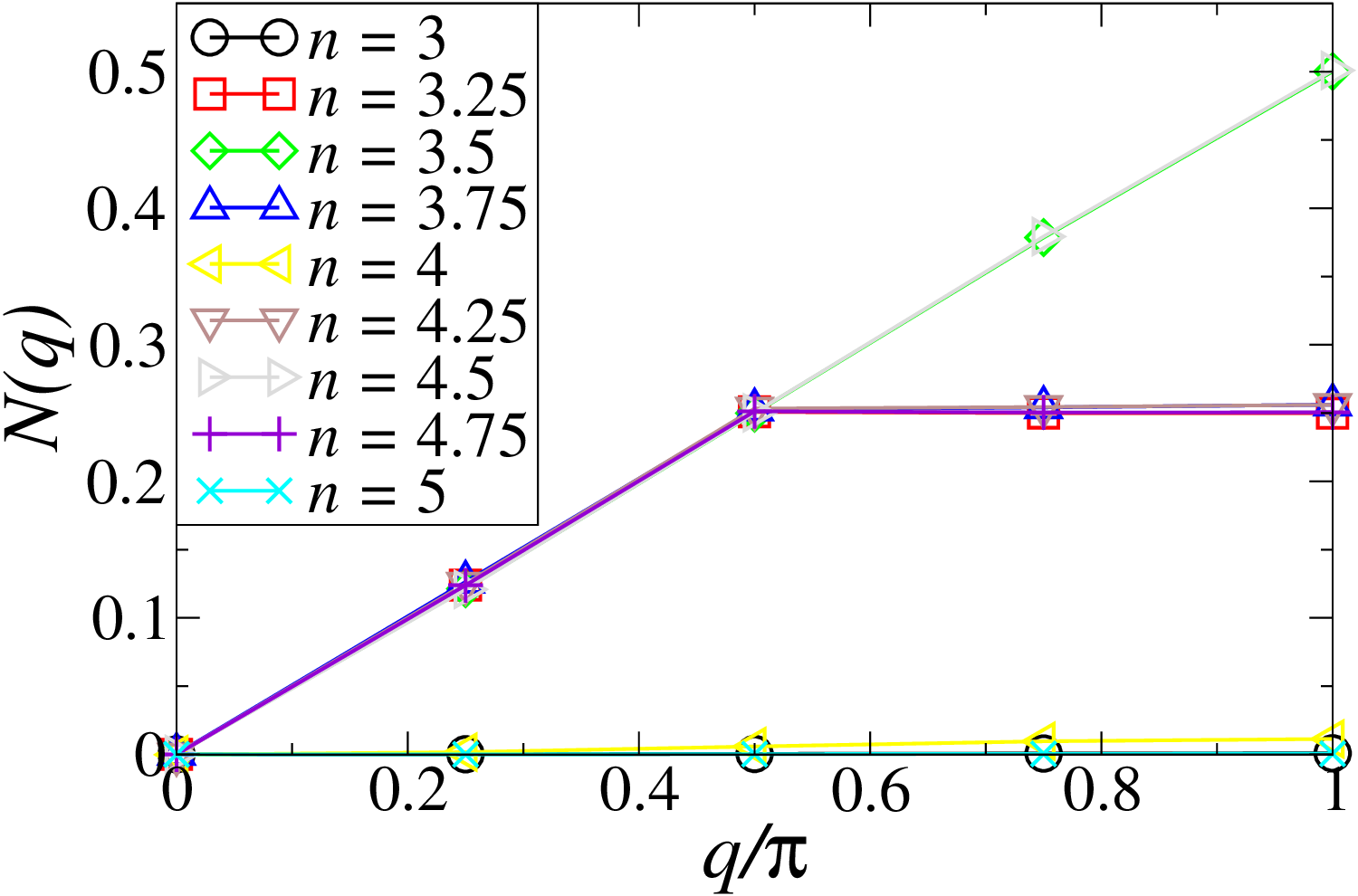}
\caption{Total-charge structure factor, $N(q)$ vs.~momentum, $q$, for several $n$ at $J/U=1/4$. The values of $U$ have been chosen such that the system is in the strong-$U$ OSMP (either OSMP2 for $n<4$ or OSMP3 for $n>4$). The effective Fermi momentum of the free spinless fermions is $n$-dependent (see main text for details).}
\label{fig:S3}
\end{figure}

Figure~\ref{fig:S3} shows the total-charge structure factor in the strong-$U$ OSMP, where spinless fermions are found. We explore the range $3\leqslant n \leqslant 5$. $N(q)$ clearly displays charge fluctuations typical of free spinless fermions. Fig.~\ref{fig:S3} shows the dependence of the effective Fermi wave vector of the spinless fermions on the total filling of the system. The $n$-dependence of the Fermi momentum is symmetric around $n=4$.

\begin{figure}[!b]
\centering
\includegraphics*[width=.8\columnwidth]{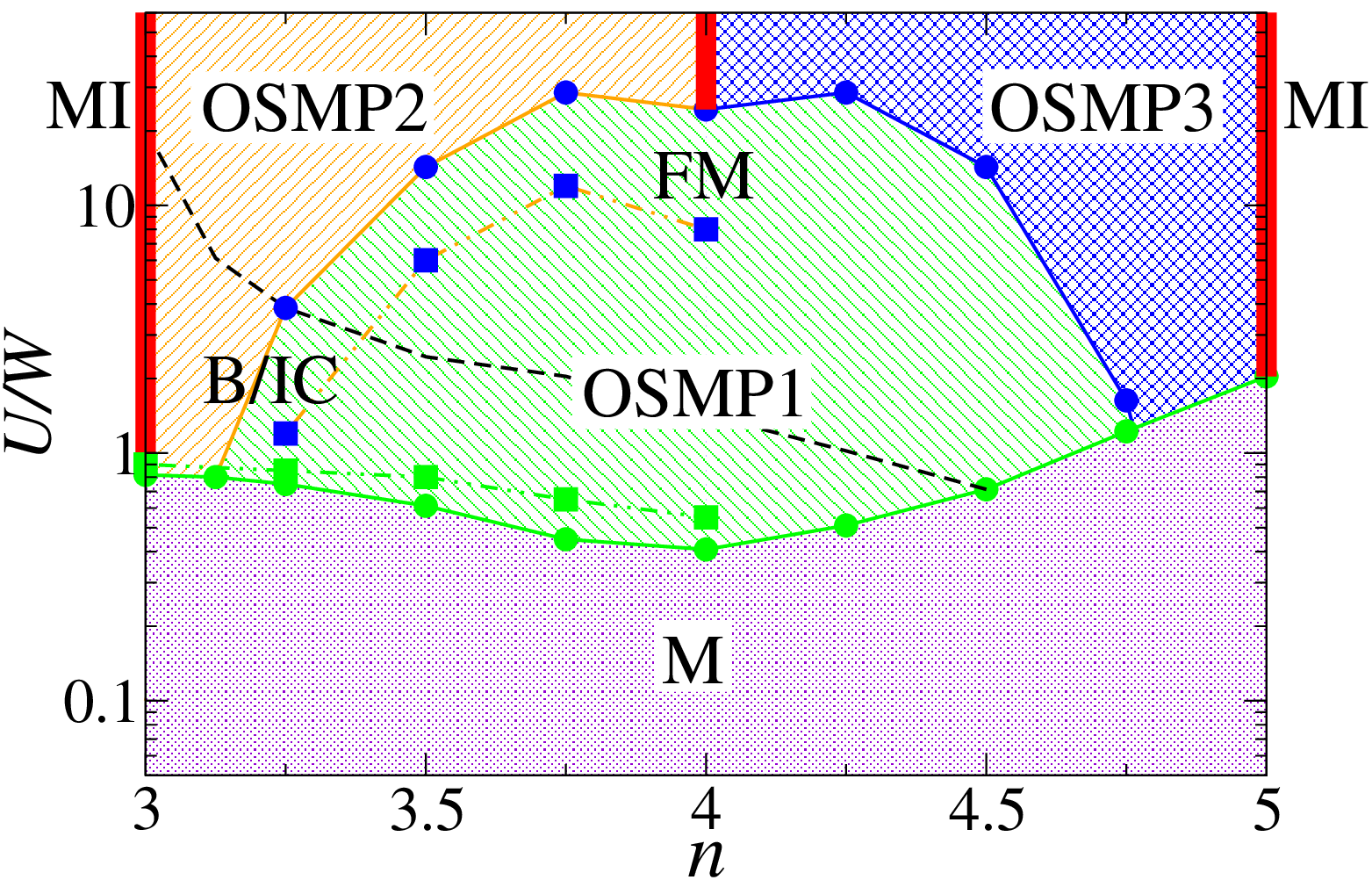}
\caption{DMRG phase diagram of the three-orbital model varying $U/W$ and $n$, at $J/U=1/4$ 
for a 48-orbital lattice. The phases are labeled as in the main text. The boundaries defined by the squares and dashed-dotted lines are for the case $J/U = 0.15$. The lines are guides to the eye.
}
\label{fig:S4}
\end{figure}

The finite-size behavior of the three-orbital model studied in this work is now discussed. Figure~\ref{fig:S4} shows the phase diagram for an 48-orbital lattice. By comparing Fig.~\ref{fig:S4} to Fig.~1, in the main text, it is possible to see that the corrections to the phase diagram due to finite-size effects are fairly small. The same phases, magnetic and charge patterns are found in both phase diagrams. More importantly, the advertized QPT between OSMP states---with different localization properties---, and all the main features of the measurements reported in the main text, are found as well in other system sizes. It is worth mentioning that the finite-size behavior exhibited by these quantities is very similar to that previously reported on similar multiorbital Hubbard models~\cite{Sakamoto02,Rincon14}. 

In addition to study the OSMP-OSMP transition for the value $J/U = 1/4$, we have also studied the presence of this transition for $J/U = 1/3$, and $0.15$. The results for the latter ratio are shown in Fig.~\ref{fig:S4}. As expected from the effects of the Hund's coupling, the boundaries separating the OSMP QPT have a lower value than in the prototypical case $J/U=1/4$, rendering the advertised transition to more realistic values. In particular, note that for $J/U = 0.15$ and electronic density $n\sim 3.25$, the critical line dividing the OSMP states is at a value of $U/W$ close to 1, namely in a realistic range of couplings. On the other hand, for the weak-coupling transition the critical values increase moderately.